\documentclass[twocolumn,10pt]{asme2ej}

\usepackage{epsfig,lipsum,amsfonts,amssymb,amsmath,graphicx,threeparttable,caption,tabularx} 

\makeatletter
\DeclareRobustCommand{\PHP}{%
	\begingroup
	\leavevmode\,\vphantom{P}%
	\dimen\z@=.5\fontcharht\font`P\relax
	\dimen\tw@=0.33333\dimen\z@
	\ooalign{%
		\raisebox{\dimexpr\dimen\z@+2\dimen\tw@-0.4pt}{\rule{\fontcharwd\font`P}{0.4pt}}\cr
		\raisebox{\dimexpr\dimen\z@+\dimen\tw@-0.2pt}{\rule{\fontcharwd\font`P}{0.4pt}}\cr
		P\cr
	}%
	\,\endgroup
}
\makeatother

\title{How much did the Tourism Industry Lost? Estimating Earning Loss of Tourism \\ in the Philippines}

\author{Raffy S. Centeno
    \affiliation{
	Senior Teacher\\
	Senior High School Department\\
	Malayan Colleges Mindanao\\
    Email: rscenteno@mcm.edu.ph
    }
}
\author{
	Judith P. Marquez
	\affiliation{
		Division Manager\\
		Planning and Monitoring Division\\
		Davao City Water District\\
		Email: jrnpmarquez\_68@yahoo.com
	}	
}

\begin{document}

\maketitle    

\begin{abstract}
{\it \indent The study aimed to forecast the total earnings lost of the tourism industry of the Philippines during the COVID-19 pandemic using seasonal autoregressive integrated moving average. Several models were considered based on the autocorrelation and partial autocorrelation graphs. Based on the Akaike's Information Criterion (AIC) and Root Mean Squared Error, ARIMA(1,1,1)$\times$(1,0,1)$_{12}$ was identified to be the better model among the others with an AIC value of $-414.51$ and RMSE of $47884.85$. Moreover, it is expected that the industry will have an estimated earning loss of around \PHP 170.5 billion pesos if the COVID-19 crisis will continue up to July. Possible recommendations to mitigate the problem includes stopping foreign tourism but allowing regions for domestic travels if the regions are confirmed to have no cases of COVID-19, assuming that every regions will follow the stringent guidelines to eliminate or prevent transmissions; or extending this to countries with no COVID-19 cases.}

\end{abstract}

\section{Introduction}

\subsection{Background of the Study}

According to the Philippine Statistics Authority, tourism accounts to 12.7\% of the country’s Gross Domestic Product in the year 2018 \cite{psa-2019-report}. Moreover, National Economic Development Authority reported that 1.5\% of the country’s GDP on 2018 is accounted to international tourism with Korea, China and USA having the largest numbers of tourists coming in \cite{}. In addition, Department of Tourism recorded that 7.4\% of the total domestic tourists or an estimated figure of 3.97 million tourists, both foreign and domestics were in Davao Region on 2018 \cite{dot-report}. Also, employment in tourism industry was roughly estimated to 5.4 million in 2018 which constitutes 13\% of the employment in the country according to the Philippine Statistics Authority \cite{psa-2018-report}.

Hence, estimating the total earnings of the tourism industry in the Philippines will be very helpful in formulating necessary interventions and strategies to mitigate the effects of the COVID-19 pandemic. This paper will serve as a baseline research to describe and estimate the earnings lost of the said industry.




\subsection{Problem Statement}

The objective of this research is to forecast the monthly earnings loss of the tourism industry during the COVID-19 pandemic by forecasting the monthly foreign visitor arrivals using Seasonal Autoregressive Integrated Moving Average. Specifically, it aims to answer the following questions:
	
	\begin{enumerate}
		\item What is the order of the seasonal autoregressive intergrated moving average for the monthly foreign visitor arrivals in the Philippines?
		
		\item How much earnings did the tourism industry lost during the COVID-19 pandemic?
	\end{enumerate}

\subsection{Scope and Limitations}

The study covers a period of approximately eight years from January 2012 to December 2019. Also, the modeling technique that was considered in this research is limited only to autoregressive integrated moving average (ARIMA) and seasonal autoregressive integrated moving average (SARIMA). Other modeling techniques were not tested and considered.

\section{Methodology}

\subsection{Research Design}

The research utilized longitudinal research design wherein the monthly foreign visitor arrivals in the Philippines is recorded and analyzed. A longitudinal research design is an observational research method in which data is gathered for the same subject repeatedly over a period of time \cite{research-design}. Forecasting method, specifically the Seasonal Autoregressive Integrated Moving Average (SARIMA), was used to forecast the future monthly foreign visitor arrivals.

In selecting the appropriate model to forecast the monthly foreign visitor arrivals in the Philippines, the Box-Jenkins methodology was used. The data set was divided into two sets: the training set which is composed of 86 data points from January 2012 to December 2018; and testing set which is composed of 12 data points from January 2019 to December 2019. The training set was used to identify the appropriate SARIMA order whereas the testing set will measure the accuracy of the selected model using root mean squared error. The best model, in the context of this paper, was characterized to have a low Akaike's Information Criterion and low root mean squared error.

\subsection{Source of Data}

The data were extracted from Department of Tourism website. The data were composed of monthly foreign visitor arrivals from January 2012 to December 2019 which is composed of 98 data points.

\subsection{Procedure for Box-Jenkins Methodology}

Box-Jenkins methodology refers to a systematic method of identifying, fitting, checking, and using SARIMA time series models. The method is appropriate for time series of medium to long length which is at least 50 observations. The Box-Jenkins approach is divided into three stages: Model Identification, Model Estimation, and Diagnostic Checking.

\begin{enumerate}
	\item \textit{Model Identification}
	
	In this stage, the first step is to check whether the data is stationary or not. If it is not, then differencing was applied to the data until it becomes stationary. Stationary series means that the value of the series fluctuates around a constant mean and variance with no seasonality over time. Plotting the sample autocorrelation function (ACF) and sample partial autocorrelation function (PACF) can be used to assess if the series is stationary or not. Also, Augmented Dickey$-$Fuller (ADF) test can be applied to check if the series is stationary or not. Next step is to check if the variance of the series is constant or not. If it is not, data transformation such as differencing and/or Box-Cox transformation (eg. logarithm and square root) may be applied. Once done, the parameters $p$ and $q$ are identified using the ACF and PACF.
	
	If there are 2 or more candidate models, the Akaike's Information Criterion (AIC) can be used to select which among the models is better. The model with the lowest AIC was selected.
	
	\item \textit{Model Estimation}
	
	In this stage, parameters are estimated by finding the values of the model coefficients which provide the best fit to the data. In this research, the combination of Conditional Sum of Squares and Maximum Likelihood estimates was used by the researcher. Conditional sum of squares was utilized to find the starting values, then maximum likelihood was applied after.
	
	\item \textit{Diagnostic Checking}
	
	Diagnostic checking performs residual analysis. This stage involves testing the assumptions of the model to identify any areas where the model is inadequate and if the corresponding residuals are uncorrelated. Box-Pierce and Ljung-Box tests may be used to test the assumptions. Once the model is a good fit, it can be used for forecasting.
	
	\item \textit{Forecast Evaluation}
	
	\hspace{5mm} Forecast evaluation involves generating forecasted values equal to the time frame of the model validation set then comparing these values to the latter. The root mean squared error was used to check the accuracy of the model. Moreover, the ACF and PACF plots were used to check if the residuals behave like white noise while the Shapiro-Wilk test was used to perform normality test.
	
\end{enumerate}

\subsection{Data Analysis}

	The following statistical tools were used in the data analysis of this study.

	\begin{enumerate}
		\item Sample Autocorrelation Function
		
			\hspace{5mm} Sample autocorrelation function measures how correlated past data points are to future values, based on how many time steps these points are separated by. Given a time series $X_t$, we define the sample autocorrelation function, $r_k$, at lag $k$ as \cite{time-series-book-01}
			\begin{equation}
				r_k = \dfrac{\displaystyle\sum_{t=1}^{N-k} (X_t - \bar{X})(X_{t+k} - \bar{X}) }{\displaystyle\sum_{t=1}^{N} (X_t - \bar{X})^2} \qquad \text{for } k = 1,2, ...  
			\end{equation}
			where $\bar{X}$ is the average of $n$ observations .

		\item Sample Partial Autocorrelation Function
		
			\hspace{5mm} Sample partial autocorrelation function measures the correlation between two points that are separated by some number of periods but with the effect of the intervening correlations removed in the series. Given a time series $X_t$, the partial autocorrelation of lag $k$ is the autocorrelation between $X_t$ and $X_{t+k}$ with the linear dependence of $X_t$ on $X_{t+1}$ through $X_{t+k-1}$ removed. The sample partial autocorrelation function is defined as \cite{time-series-book-01}
			\begin{equation}
				\phi_{kk} = \dfrac{r_k - \displaystyle\sum_{j = 1}^{h-1} \phi_{k-1,j} r_{k-j}}{1 - \displaystyle\sum_{j = 1}^{j - 1} \phi_{k-1,j} r_j }
			\end{equation}
			where  $\phi_{k,j} = \phi_{k-1,j} - \phi_{k,k} \phi_{k-1,k-j}, \text{for } j = 1,2, ..., k-1$, and $r_k$ is the sample autocorrelation at lag $k$. 
		
		\item Root Mean Square Error (RMSE)
		
			\hspace{5mm} RMSE is a frequently used measure of the difference between values predicted by a model and the values actually observed from the environment that is being modelled. These individual differences are also called residuals, and the RMSE serves to aggregate them into a single measure of predictive power. The RMSE of a model prediction with respect to the estimated variable $X_{\text{model}}$ is defined as the square root of the mean squared error \cite{LSTM-book-01}
			
			\begin{center}
				$RMSE = \sqrt{\dfrac{1}{n}\displaystyle\sum_{i=1}^{n} (\hat{y}_{i} - y_i)^2}$
			\end{center}
			
			where $\hat{y_i}$ is the predicted values, $y_i$ is the actual value, and $n$ is the number of observations.

		\item Akaike's Information Criterion (AIC)
		
			\hspace{5mm} The AIC is a measure of how well a model fits a dataset, penalizing models that are so flexible that they would also fit unrelated datasets just as well. The general form for calculating the AIC is \cite{time-series-book-01}
			\begin{equation}
				AIC_{p,q} = \dfrac{-2 \ln(\text{maximized likelihood}) + 2r}{n}
			\end{equation}			
			where $n$ is the sample size, $r = p + q + 1$ is the number of estimated parameters, and including a constant term.
		
		\item Ljung$-$Box Q* Test
		
			\hspace{5mm} The Ljung$-$Box statistic, also called the modified Box-Pierce statistic, is a function of the accumulated sample autocorrelation, $r_j$, up to any specified time lag $m$. This statistic is used to test whether the residuals of a series of observations over time are random and independent. The null hypothesis is that the model does not exhibit lack of fit and the alternative hypothesis is the model exhibits lack of fit. The test statistic is defined as \cite{time-series-book-01}			
			\begin{equation}
				Q^* = n (n+2) \displaystyle\sum_{k = 1}^{m} \dfrac{ \hat{r}^2_k }{n - k}
			\end{equation}			
			where $\hat{r}^2_k$ is the estimated autocorrelation of the series at lag $k$, $m$ is the number of lags being tested, $n$ is the sample size, and the given statistic is approximately Chi Square distributed with $h$ degrees of freedom, where $h = m - p - q$.

		\item Conditional Sum of Squares
		
			\hspace{5mm} Conditional sum of squares was utilized to find the starting values in estimating the parameters of the SARIMA process. The formula is given by \cite{forecast}			
			\begin{equation}
				\hat{\theta}_n = \arg \min\limits_{\theta \in \ominus} s_n (\theta)
			\end{equation}			
			where $s_n(\theta) = \dfrac{1}{n}\displaystyle\sum_{t=1}^{n}e^2_t(\theta) \ , e_t(\theta) = \displaystyle\sum_{j=0}^{t-1} \alpha_j(\theta)x_{t-j}$, and $\ominus \subset \mathbb{R}^p$ is a compact set.
		
		\item Maximum Likelihood
		
			\hspace{5mm} According to \cite{forecast}, once the model order has been identified, maximum likelihood was used to estimate the parameters $c$, $\phi_1, ..., \phi_p, \theta_1, ..., \theta_q$. This method finds the values of the parameters which maximize the probability of getting the data that has been observed . For SARIMA models, the process is very similar to the least squares estimates that would be obtained by minimizing
			\begin{equation}
				\displaystyle\sum_{t=1}^{T} \epsilon^2_t
			\end{equation}
			where $\epsilon_t$ is the error term.
			
		\item Box$-$Cox Transformation
		
			\hspace{5mm} Box$-$Cox Transformation is applied to stabilize the variance of a time series. It is a family of transformations that includes logarithms and power transformation which depend on the parameter $\lambda$ and are defined as follows \cite{Daimon2011}
			
			\begin{center}
				$y^{(\lambda)}_i = 
				\begin{cases}
				\dfrac{y^\lambda_i - 1}{\lambda}	&	\text{, if } \lambda \neq 0\\
				\ln y_i							&	\text{, if } \lambda = 0
				\end{cases}
				$
				$
				\qquad \qquad
				$
				$
				w_i =
				\begin{cases}
				y_i^{\lambda}	& \text{, if } \lambda \neq 0\\
				\ln y_i			& \text{, if } \lambda = 0
				\end{cases}
				$
			\end{center}
			
			where $y_i$ is the original time series values, $w_i$ is the transformed time series values using Box-Cox, and $\lambda$ is the parameter for the transformation.

	\end{enumerate}

\subsection{Statistical Software}

R is a programming language and free software environment for statistical computing and graphics that is supported by the R Foundation for Statistical Computing \cite{R-software}. R includes linear and nonlinear modeling, classical statistical tests, time-series analysis, classification modeling, clustering, etc. The `forecast' package \cite{forecast} was utilized to generate time series plots, autocorrelation function/partial autocorrelation function plots, and forecasting. Also, the `tseries' package \cite{tseries} was used to perform Augmented Dickey-Fuller (ADF) to test stationarity. Moreover, the `lmtest' package \cite{lmtest} was used to test the parameters of the SARIMA model. Finally, the `ggplot2' \cite{ggplot2}, `tidyr' \cite{tidyr}, and `dplyr' \cite{dplyr} were used to plot time series data considered during the conduct of the research.

\section{Results and Discussion}

\begin{figure}[h]
	\includegraphics[width=3.4in]{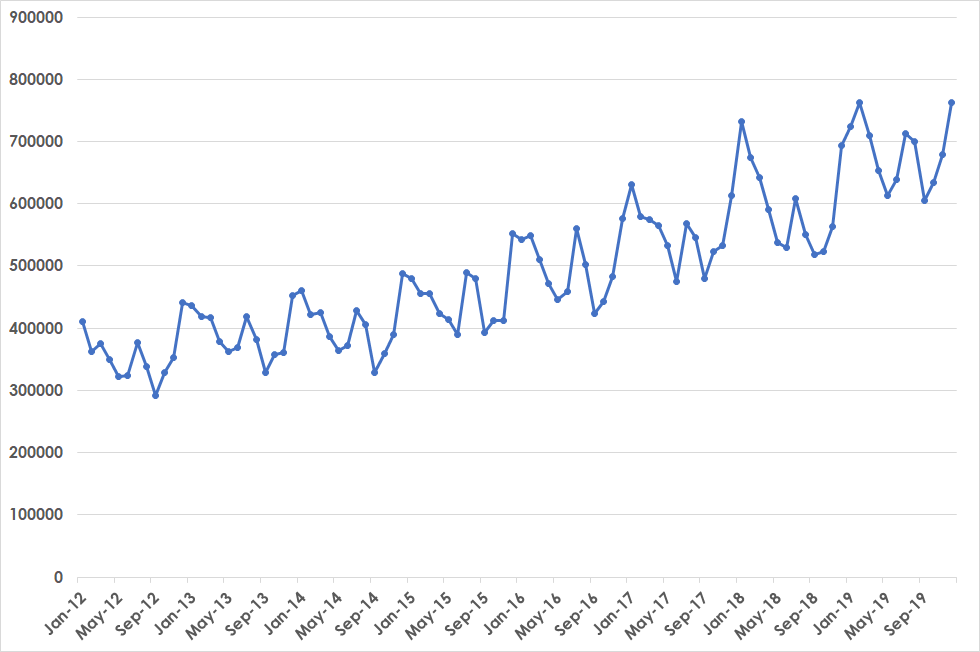}
	\caption{Monthly Foreign Visitor Arrivals}
	\label{rd01}
\end{figure}

Line plot was used to describe the behavior of the monthly foreign visitor arrivals in the Philippines. Figure~\ref{rd01} shows that there is an increasing trend and a seasonality pattern in the time series. Specifically, there is a seasonal increase in monthly foreign visitor arrivals every December and a seasonal decrease every September. These patterns suggest a seasonal autoregressive integrated moving average (SARIMA) approach in modeling and forecasting the monthly foreign visitor arrivals in the Philippines.

\begin{table}[h]
	\captionof{table}{AIC and RMSE of the Two Models Considered}
	\label{rd02}
	\renewcommand{\arraystretch}{1}
	\begin{tabularx}{3.35in}{Xcc}	\hline
		\textbf{Model} 						&	\textbf{AIC}	&	\textbf{RMSE}	\\	\hline
		ARIMA (0,1,2)$\times$(1,0,1)$_{12}$	&	$-414.56$		&	49517.48		\\
		ARIMA (1,1,1)$\times$(1,0,1)$_{12}$	&	$-414.51$		&	47884.85		\\	 \hline
	\end{tabularx}
\end{table}

Akaike Information Criterion and Root Mean Squared Error were used to identify which model was used to model and forecast the monthly foreign visitor arrivals in the Philippines. Table~\ref{rd02} shows the top two SARIMA models based on AIC generated using R. ARIMA (1,1,1)$\times$(1,0,1)$_{12}$ has the lowest AIC with a value of $-414.56$ which is followed by ARIMA (1,1,1)$\times$(1,0,1)$_{12}$ with an AIC value of $-414.51$. Model estimation was performed on both models and generated significant parameters for both models (refer to Appendix A.2). Moreover, diagnostic checking was performed to assess the model. Both models passed the checks using residual versus time plot, residual versus fitted plot, normal Q-Q plot, ACF graph, PACF graphs, Ljung-Box test, and Shapiro-Wilk test (refer to Appendix A.3). Finally, forecast evaluation was performed to measure the accuracy of the model using an out-of-sample data set (refer to Appendix A.4). ARIMA (1,1,1)$\times$(1,0,1)$_{12}$ produced the lowest RMSE relative to ARIMA (0,1,2)$\times$(1,0,1)$_{12}$. Hence, the former was used to forecast the monthly foreign visitor arrivals in the Philippines.

\subsection{How much Foreign Tourism Earnings was Lost during the COVID-19 Pandemic Crisis}

\begin{figure}[h]
	\includegraphics[width=3.4in]{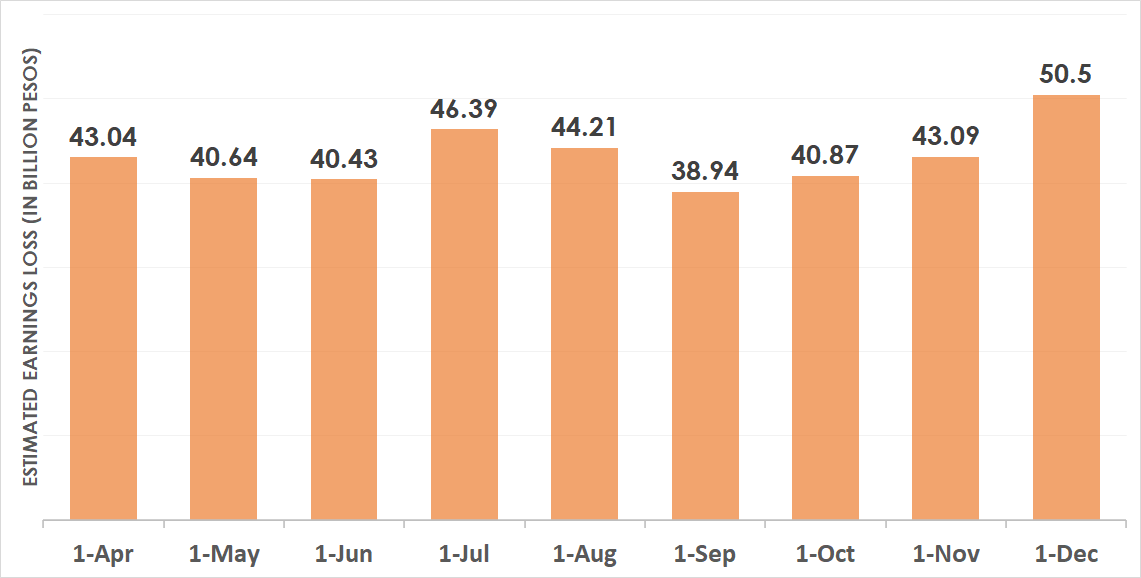}
	\caption{Expected Monthly Earnings Loss}
	\label{rd03}
\end{figure}

Figure~\ref{rd03} shows the estimated earnings loss (in billion pesos) of the tourism industry of the Philippines every month from April 2020 to December 2020. According to the Department of Tourism, the Average Daily Expenditure (ADE) for the month in review is \PHP 8,423.98 and the Average Length of Stay (ALoS) of tourists in the country is recorded at 7.11 nights. The figures were generated by multiplying the forecasted monthly foreign visitor arrivals, ADE, and ALoS (rounded to 7) \cite{dot-report}. Moreover, it is forecasted under community quarantine that the recovery time will take around four to five months (up to July) \cite{forecast-covid}. With this, the estimated earning loss of the country in terms of tourism will be around 170.5 billion pesos.

\section{Conclusions and Recommendations}

\subsection{Conclusions}

	Based on the results presented on the study, the following findings were drawn:
	\begin{enumerate}
		\item The order of SARIMA model used to forecast the monthly foreign visitor arrival is ARIMA (1,1,1)$\times$(1,0,1)$_{12}$ since it produced a relatively low AIC of $-414.51$ and the lowest RMSE of 47884.85 using an out-of-sample data. This means that the model is relatively better among other SARIMA models considered in forecasting the monthly foreign visitor arrivals in the Philippines.
		
		\item If the COVID-19 Pandemic lasts up to five months, the tourism industry of the Philippines will have an estimated earnings loss of about \PHP 170.5 billion. Assumptions about average daily expenditure and average length of stay of tourists were based on the Department of Tourism reports.
	\end{enumerate}

\subsection{Recommendations}

The projected \PHP 170.5 billion loss on Philippine’s foreign tourism is really a huge money. Regaining such loss the soonest time, however, would only jeopardize the lives of the Filipino people. On the other hand, the government can, perhaps, reopen the Philippines’ domestic tourism. This would somehow help regain the country’s loss on revenue from tourism, although not fully.
 
However, the following recommendations, shown in scenarios/options below, may be helpful in regaining it, both in foreign and domestic tourism, and ensuring safety among Filipinos, as well.

	\begin{enumerate}
		\item Option 1: Stop foreign tourism until the availability of the vaccine, but gradually open domestic tourism starting July of 2020. In this scenario/option, the following considerations may be adhered to, viz.
			
			\begin{enumerate}
				\item not all domestic tourism shall be reopened in the entire country; only those areas with zero covid-19 cases; 
				\item for areas where domestic tourism is allowed/reopened, appropriate guidelines should be strictly implemented by concerned departments/agencies to eliminate/prevent covid-19 transmission; and
				\item digital code that would help in tracing the contacts and whereabouts of domestic tourists, as being used in China and Singapore, should be installed before the reopening of the domestic tourism.				
			\end{enumerate}
		
		\item Option 2: Gradual opening of foreign tourism starting July 2020 and full reopening of domestic tourism on the first semester of 2021 or when the covid-19 cases in the Philippines is already zero. However, the following considerations should be satisfied, viz.
		
			\begin{enumerate}
				\item only countries with covid-19 zero cases are allowed to enter the Philippines; 
				\item appropriate guidelines should be strictly implemented by concerned departments/ agencies both for foreign and domestic tourism to eliminate/ prevent the spread of the said virus; and
				\item digital code that would help in tracing the contacts and whereabouts of foreign tourists, as being used in China and Singapore, should be installed before reopening the foreign tourism in the Philippines. 				
			\end{enumerate}			
	\end{enumerate}

%
% ...

\bibliographystyle{asmems4}

\bibliography{asme2e}

\appendix       

\section{Appendices}

\subsection{Model Identification}

\begin{figure}[h!]
	\includegraphics[width=3.4in]{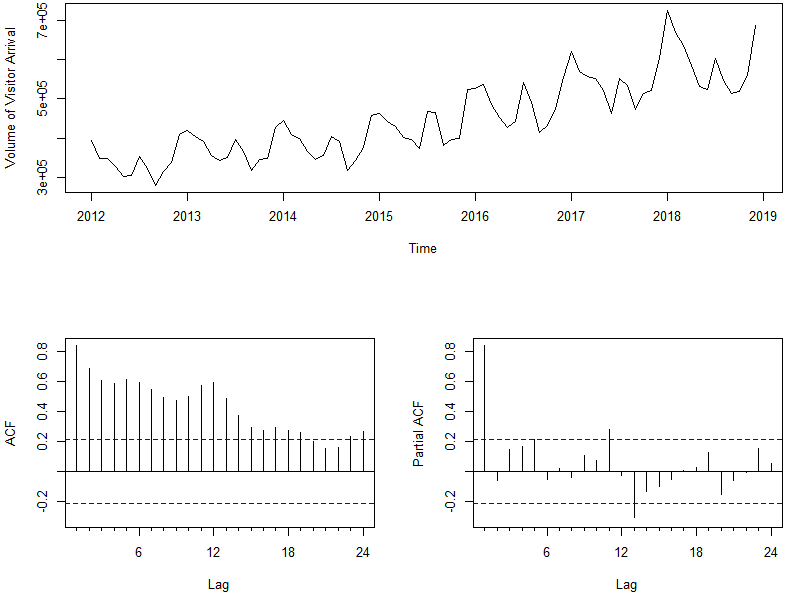}
	\caption{Line, ACF, and PACF Plot of Monthly Visitor Arrivals}
	\label{mi01}
\end{figure}

Line, ACF, and PACF graph were used to identify the model to be used to forecast the monthly visitor arrivals in the Philippines. The line graph in Figure~\ref{mi01} shows an increasing trend which suggests a non-stationary behavior. This is supported by the ACF and PACF plots which shows a slow decay in all the lags of the former and the the first lag is significant for the latter. Moreover, the line plot slightly display an increasing variance across time. Therefore, data transformation such as differencing and Box-Cox transform were applied to the time series data.

\begin{figure}[h!]
	\includegraphics[width=3.4in]{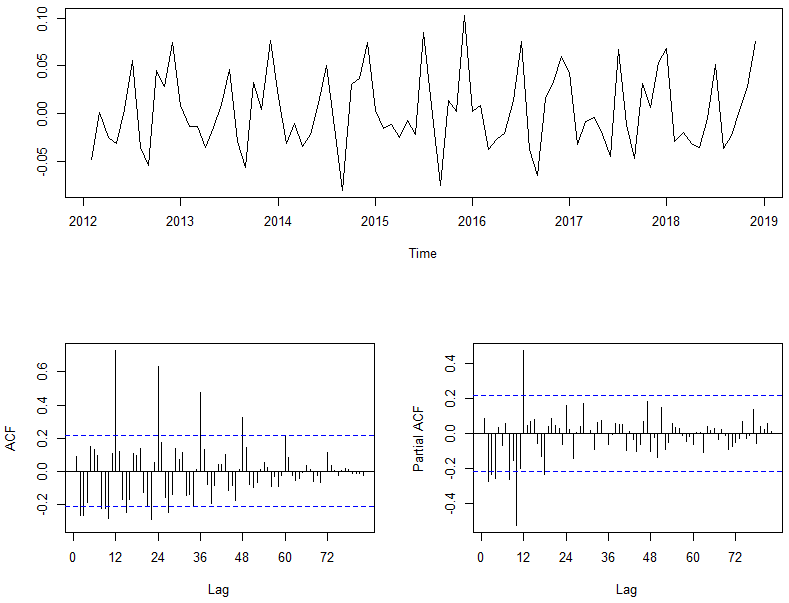}
	\caption{Line, ACF, and PACF Plot of the Transformed Data}
	\label{mi02}
\end{figure}

Figure~\ref{mi02} show the line, ACF, and PACF graph of the transformed data. The line graph shows that the data is stationary which is supported by both ACF and PACF graphs. Moreover, ACF graph suggests a seasonal pattern in the data since 12$^{\text{th}}$, 24$^{\text{th}}$, 36$^{\text{th}}$, 48$^{\text{th}}$, and 60$^{\text{th}}$ lags are significant. This is also true in the case of PACF since the 12$^{\text{th}}$ lag is significant.

\begin{table}[h]
	\captionof{table}{Akaike's Information Criterion of each ARIMA Model}
	\label{mi03}
	\renewcommand{\arraystretch}{1.2}
	\begin{tabularx}{3.35in}{Xc}	\hline
		\textbf{Model} 								&	\textbf{AIC}	\\	\hline
		ARIMA (0,1,2)$\times$(1,0,1)$_{12}$		&	$-414.56$		\\
		ARIMA (1,1,1)$\times$(1,0,1)$_{12}$		&	$-414.51$		\\	\hline
	\end{tabularx}
\end{table}

Akaike's Information Criterion (AIC) was used to identify the best SARIMA model from among the models considered. Table~\ref{mi03} shows the top 2 models with the least AIC, namely: ARIMA (0,1,2)$\times$(1,0,1)$_{12}$ (Model 1) and ARIMA (1,1,1)$\times$(1,0,1)$_{12}$ (Model 2)with an AIC of $-414.56$ and $-414.51$, respectively.

\subsection{Model Estimation}

\begin{table}[h]
	\captionof{table}{Model Estimation of Model 1}
	\label{mi04}
	\renewcommand{\arraystretch}{1.2}
	\begin{tabularx}{3.35in}{Xrcrl} \hline
		\textbf{Variable}	& $\beta$	&Std. Error	&	z-value	&					\\ \hline
		ma1					&$-0.502$	&0.108	 	&$-4.662$	&\hspace{-4mm}***	\\
		ma2					&$-0.224$	&0.104	 	&$-2.154$	&\hspace{-4mm}*	\\
		sar1				&0.993		&0.011	 	&87.960		&\hspace{-4mm}***	\\
		sma1				&$-0.701$	&0.206	 	&$-3.408$	&\hspace{-4mm}***	\\
		log-likelihood		&			&			&	212.28	&					\\
		$\sigma^2$			&			&			&	0.0002	&					\\	
		AIC					&			&			&$-414.56$	&					\\	 \hline
	\end{tabularx}
	*** p $< 0.001$, * p $<0.05$
\end{table}
	
The combination of conditional sum of squares and maximum likelihood estimates were used to estimate the parameters of the three moving averages and test its significance. Table~\ref{mi04} shows the estimated coefficients, standard errors, z-values, and p-values of each parameters of ARIMA (0,1,2)$\times$(1,0,1)$_{12}$. Since the p-value of the parameter is less than 0.05, there is sufficient evidence to say that the estimate of the moving averages, seasonal autoregressive, and seasonal moving average are significantly different from zero.
	
\begin{table}[h]
	\captionof{table}{Model Estimation of Model 2}
	\label{mi05}
	\renewcommand{\arraystretch}{1.2}
	\begin{tabularx}{3.35in}{Xrcrl} \hline
		\textbf{Variable}	& $\beta$	&Std. Error	&	z-value	&					\\ \hline
		ar1					&$0.328$	&0.153	 	&$2.141$	&\hspace{-4mm}*	\\
		ma1					&$-0.830$	&0.086	 	&$-9.697$	&\hspace{-4mm}***	\\
		sar1				&0.992		&0.011	 	&86.543		&\hspace{-4mm}***	\\
		sma1				&$-0.701$	&0.207	 	&$-3.388$	&\hspace{-4mm}***	\\
		log-likelihood		&			&			&	212.25	&					\\
		$\sigma^2$			&			&			&	0.0002	&					\\	
		AIC					&			&			&$-414.51$	&					\\	 \hline
	\end{tabularx}
	*** p $< 0.001$, * p $<0.05$
\end{table}
	
The combination of conditional sum of squares and maximum likelihood estimates were used to estimate the parameters of the three moving averages and test its significance. Table~\ref{mi05} shows the estimated coefficients, standard errors, z-values, and p-values of each parameters of ARIMA (1,1,1)$\times$(1,0,1)$_{12}$. Since the p-value of the parameter is less than 0.05, there is sufficient evidence to say that the estimate of the moving averages, seasonal autoregressive, and seasonal moving average are significantly different from zero.

\subsection{Diagnostic Checking}

\begin{figure}[h!]
	\includegraphics[width=3.4in]{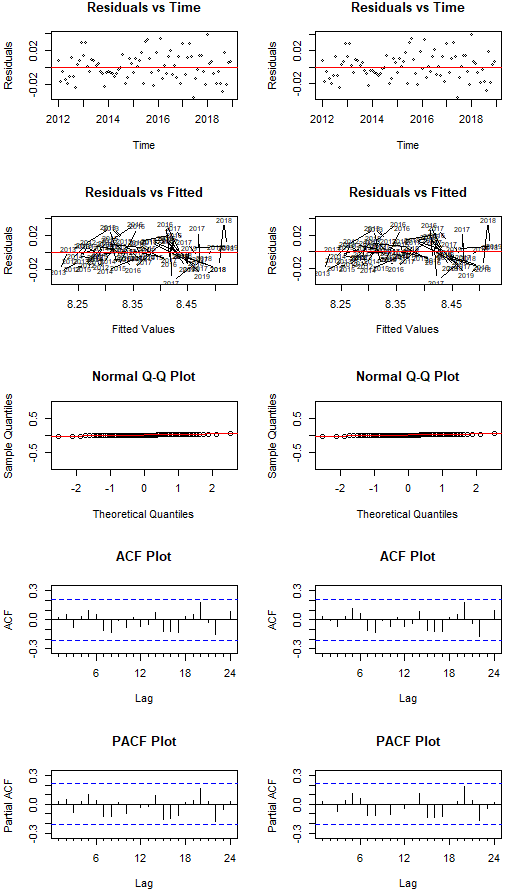}
	\caption{Line, ACF, and PACF Plot of the Two Models}
	\label{mi06}
\end{figure}

Residual versus Time, Residual versus Fitted, and Normal Q-Q Plot were used to perform diagnostic checking for the model whereas ACF and PACF plots of the residuals were used to check if there are remaining patterns that should be accounted by the model 1 and 2. Graphs for Model 1 are displayed in the left whereas graphs for Model 2 are displayed in the right. Figure~\ref{mi06} shows that the residuals and time does not display correlation between the two variables. Therefore, this scatter plot suggests that the residuals has no serial correlation, that is, there is no interdependence between time and residuals. This is supported by Ljung-Box test which suggests that the error terms behave randomly for both Model 1 (Q$(20)=20.109$, df=20, Model df=4, p$=0.4511$) and Model 2 (Q$(20)=20.941$, df=20, Model df=4, p$=0.4006$). In addition, the residuals versus fitted values scatter plot displays no visible funneling pattern which indicates that the variances of the error term are relatively equal. Moreover, the normal Q-Q plot suggests that the residuals are normally distributed since most of the the values lie along a line. This is supported by Shapiro-Wilk test which suggest that the error is normally distributed for both Model 1 (W$=0.98062$, p$=0.2372$) and Model 2 (W$=0.9852$, p$=0.4513$). Finally, ACF and PACF graphs displays that all of the lags are within the acceptable limits. Therefore, all the lags are not significant which means that the residuals of the model may be considered as white noise. Hence, the residuals are assumed to be Guassian white noise.

\subsection{Forecast Evaluation}

\begin{figure}[h!]
	\includegraphics[width=3.4in]{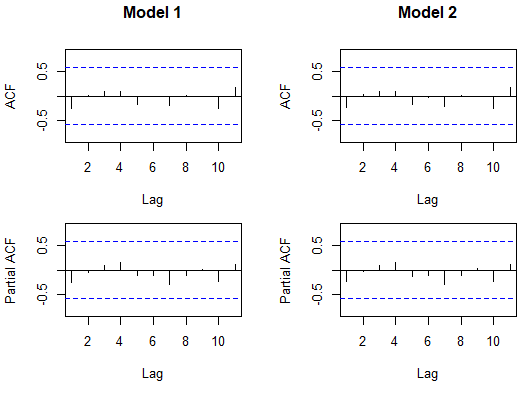}
	\caption{Line, ACF, and PACF Plot of the Two Models}
	\label{mi07}
\end{figure}

Figure~\ref{mi07} shows the ACF and PACF graphs of the forecast errors both models. All of the autocorrelation and partial autocorrelation are  within the limits which means that these values are not significant. Therefore, the forecast errors are considered white noise.

\begin{table}[h]
	\captionof{table}{Ljung-Box and Kolmogorov-Smirnov Test}
	\label{mi08}
	\renewcommand{\arraystretch}{1}
	\begin{tabularx}{3.35in}{Xcc}	\hline
		\textbf{Statistic} 	&	\textbf{Model 1}	&	\textbf{Model 2}	\\	\hline
		Shapiro-Wilk		&	$0.878$				&	$0.874$				\\
		RMSE				&	49517.48			&	47884.85		\\	 \hline
	\end{tabularx}
\end{table}

Shapiro-Wilk test was used to test if the forecast errors of both models were normally distributed. The results show that there is no sufficient evidence to say that the forecast error terms are not normally distributed for both Model 1 (W$=0.0.878$, p$=0.082$) and Model 2 (W$=0.875$, p$=0.075$). This means that it can be assumed that the forecast errors are normally distributed. Moreover, root mean squared error was used to identify which model has better forecast accuracy. The results show that Model 2 has the lowest RMSE which means that Model 2 is relatively accurate compared to Model 1.

Based on the diagnostic presented, the models satisfied all the assumptions of a seasonal autoregressive integrated moving average model. Furthermore, Model 2 is relatively accurate compared to Model 1 based on the RMSE of each model. Hence, the ARIMA (1,1,1)$\times$(1,0,1)$_{12}$ was used to forecast the monthly visitor arrivals in the Philippines.

\end{document}